\documentclass[jkps,preprint,fleqn,showpacs,showkeys]{revtex4}
\usepackage{graphicx}
\usepackage{amssymb}
\usepackage{amsmath}
\usepackage{bm}
\begin{document}
\setcounter{page}{0}
\title[]{Nontrivial ferrimagnetism of the Heisenberg model on the Union Jack strip lattice}
\author{Tokuro \surname{Shimokawa}}
\email{t.shimokaw@gmail.com}
\author{Hiroki \surname{Nakano}}

\affiliation{Graduate School of Material Science, University of Hyogo, Kamigori, Hyogo 678-1297, Japan}

\date[]{Received 31 May 2012}

\begin{abstract}
We study the ground-state properties of the $S=1/2$ antiferromagnetic Heisenberg model on the Union Jack strip lattice by using the exact-diagonalization and density matrix renormalization group methods.
We confirm a region of the intermediate-magnetization state between the N$\rm{\acute e}$el-like spin liquid state and the conventional ferrimagnetic state of Lieb-Mattis type.
In the intermediate-state, we find that the spontaneous magnetization changes gradually with respect to the strength of the inner interaction.
In addition, the local magnetization clearly shows an incommensurate modulation with long-distance periodicity in the intermediate-magnetization state.
These characteristic behaviors lead to the conclusion that the intermediate-magnetization state is the non-Lieb-Mattis ferrimagnetic one.
We also discuss the relationship between the ground-state properties 
of the $S=1/2$ antiferromagnetic Heisenberg model on the original Union Jack lattice and those on our strip lattice.

\end{abstract}



\maketitle

\section{INTRODUCTION}
Ferrimagnetism is a fundamental phenomenon in the field of magnetism. 
The most famous type of ferrimagnetism is called Lieb-Mattis (LM) one\cite{Lieb, Marshall, Takano, Okamoto, Tonegawa, Sakai}.
For example, this ferrimagnetism appears in the ground state of the ($s$, $S$)=(1/2, 1) mixed spin chain with nearest-neighbor antiferromagnetic interaction.
In this system, the occurrence of the LM ferrimagnetism originates from the situation that two different spins are arranged alternately in a line owing to the AF interaction.
In the LM ferrimagnetic state, the spontaneous magnetization occurs and the magnitude is fixed to a simple fraction of the saturated magnetization. 
As in the case of this mixed spin chain, not only the magnetic properties but also the occurrence mechanism of the LM ferrimagnetism are well known 
since this type of ferrimagnetism has been studied extensively.
Especially, the ferrimagnetism in the quantum Heisenberg spin model on the bipartite lattice without frustration is well understood within 
the Marshall-Lieb-Mattis (MLM) theorem\cite{Lieb, Marshall}. 

On the other hand, a new type of ferrimagnetism that is clearly different from the LM ferrimagnetism has been found in the ground state 
of several one-dimensional frustrated Heisenberg spin systems\cite{PF1, PF2, PF3, PF4, PF5, Shimokawa, Shimokawa2}. 
The spontaneous magnetization in this new type of ferrimagnetism 
changes gradually with respect to the strength of frustration. 
In addition, the incommensurate modulation with long-distance periodicity in local magnetizations 
is observed as a characteristic quantum behavior of the new type of ferrimagnetism.
Hereafter, we call the new type of ferrimagnetism non-Lieb-Mattis (NLM) type. 
The mechanism of the occurrence of the NLM ferrimagnetism have not yet been clarified in contrast to the case of the LM ferrimagnetism.

\begin{figure}
\includegraphics[width=16.0cm]{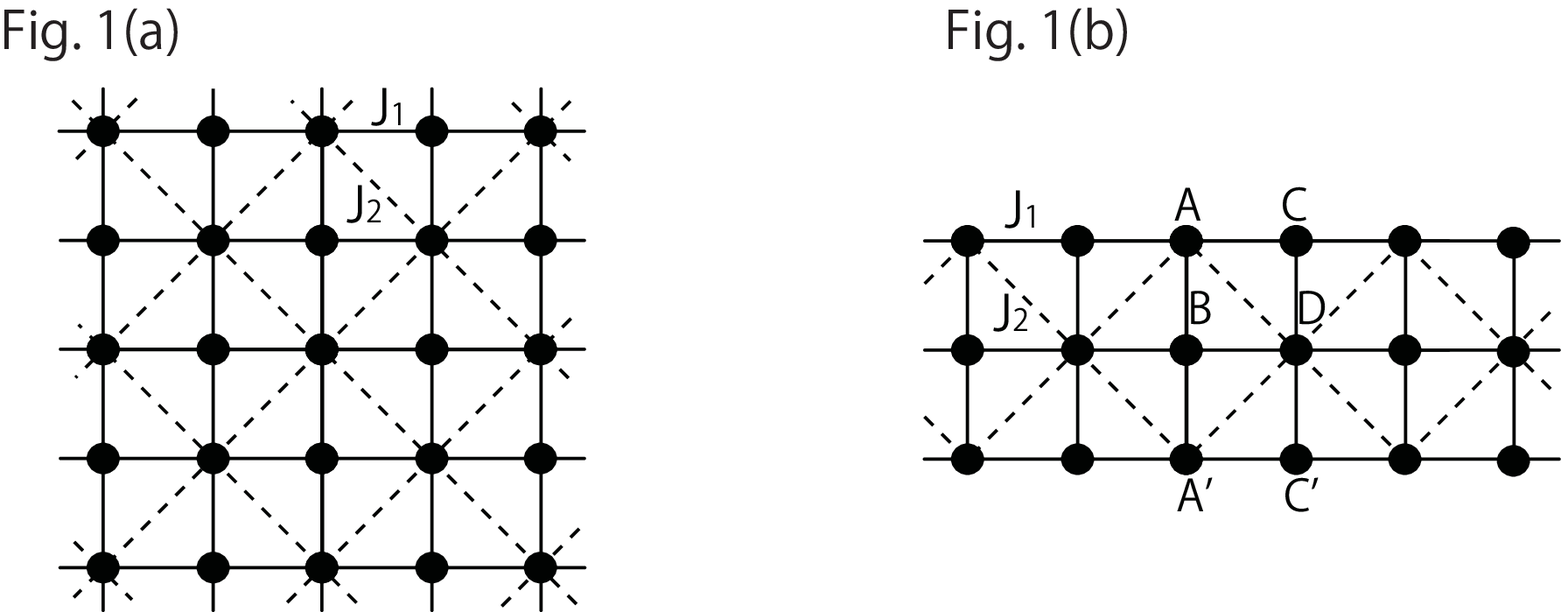}
\caption{(Color online) Structures of the lattices: the Union Jack lattice (a), the Union Jack strip lattice (b).
An $S=1/2$ spin is located at each site denoted by a black circle.
Antiferromagnetic bonds $J_{1}$ (bold straight line) 
and $J_{2}$ (dashed line) are represented. 
Sublattices in a unit cell of lattice (b) are represented 
by A, ${\rm A}^{\prime}$, B,  C, ${\rm C}^{\prime}$, and D.
}\label{fig1}
\end{figure}

\begin{figure}
\includegraphics[width=16.0cm]{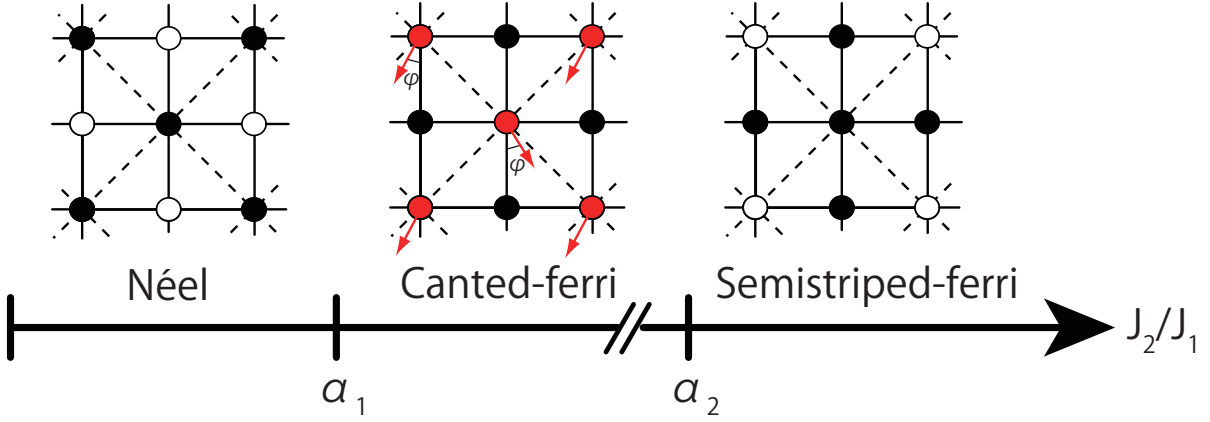}
\caption{ (Color online) Ground-state phase diagram of the $S=1/2$ Heisenberg model on the Union Jack lattice depicted Fig. \ref{fig1}(a).  
Here, black and white circles represent up-spin and down-spin respectively.}
\label{fig2}
\end{figure}

Historically, some candidates of the NLM ferrimagnetism among the 2D systems were already reported. 
For examples, there are the mixed-spin $J_{1}$-$J_{2}$ Heisenberg model on the square lattice\cite{J1J2-square} and the $S=1/2$ Heisenberg model on the Union Jack lattice  of Fig. \ref{fig1}(a) \cite{UJack1, UJack2, UJack3, UJack4}. 
These 2D frustrated systems have the intermediate ground-state, namely ``canted-ferrimagnetic state'' as described in Fig. \ref{fig2}, in which the spontaneous magnetization is changed when the inner interaction of the system is varied.
It has not been, however, investigated whether the incommensurate modulation with long-distance periodicity exists or not in the local magnetization of the intermediate-magnetization state owing to the difficulty of treating these 2D frustrated systems numerically and theoretically. Therefore, the relationships between the intermediate-magnetization states of these 2D frustrated systems and the NLM ferrimagnetic state are still unclear.

Under such circumstances, quite recently,
the $S=1/2$ antiferromagnetic Heisenberg model
on the spatially anisotropic kagome
lattice was studied\cite{Nakano-2D}.
In this model, the intermediate-magnetization states exist
between the LM ferrimagnetic state and
the nonmagnetic one\cite{ISO-KGM1, ISO-KGM2, ISO-KGM3, ISO-KGM4, ISO-KGM5}.
It was reported that the local magnetization
in these intermediate-state
shows large dependence on the position of the sites
although it is difficult to judge clearly
whether the incommensurate modulation
with long-distance periodicity is present or absent.
In addition, the $S=1/2$ Heisenberg models on the quasi-one-dimensional
kagome strip lattices were studied\cite{KGM-strip1, KGM-strip2}.
These strip lattices share the same lattice structure
in their inner part with the spatially anisotropic kagome lattice.
The local magnetizations in the intermediate-state 
clearly show incommensurate modulations
with long-distance periodicity irrespective of the strip width.
Therefore, these results strongly suggest that
the intermediate-magnetization states
not only of the kagome strip lattices
but also of the original kagome lattice are the NLM ferrimagnetism.


These kagome results motivate us to investigate the ground-state properties of the quasi-one-dimensional strip model whose lattice structure is common to the part of the 2D lattice known as the other candidates of the NLM ferrimagnetism. 
In this study, we treat the $S=1/2$ Heisenberg model on the Union Jack strip lattice depicted in Fig. \ref{fig1}(b).
This strip lattice share the same lattice structure in the inner part with the original Union Jack lattice depicted in Fig. \ref{fig1}(a).
Our numerical calculations lead to the conclusion that the NLM ferrimagnetic phase appears in the ground-state of the Union Jack strip model of Fig. \ref{fig1}(b).
We also discuss the relationship between the ground-state properties of the present strip model and those of the original 2D model.

\section{Model}
The Hamiltonian of the $S=1/2$ antiferromagnetic Heisenberg model on the Union Jack strip lattice depicted in Fig. \ref{fig1}(b) is given by 
\begin{eqnarray}
\label{Hamiltonian1}
\mathcal{H} &=&
  J_{1} \sum_{i} [  {\bf S}_{i,{\rm A}}\cdot {\bf S}_{i,{\rm C}}
                     + {\bf S}_{i,{\rm B}}\cdot {\bf S}_{i,{\rm D}}
                     + {\bf S}_{i,{\rm A}^{\prime}}\cdot {\bf S}_{i,{\rm C}^{\prime}}  
                     + {\bf S}_{i,{\rm A}}\cdot {\bf S}_{i,{\rm B}}
                     + {\bf S}_{i,{\rm B}}\cdot {\bf S}_{i,{\rm A}^{\prime}} 
\nonumber \\                     
                     &+& {\bf S}_{i,{\rm C}}\cdot {\bf S}_{i,{\rm D}}
                     + {\bf S}_{i,{\rm D}}\cdot {\bf S}_{i,{\rm C}^{\prime}}
                     + {\bf S}_{i,{\rm C}}\cdot {\bf S}_{i+1,{\rm A}}
                     + {\bf S}_{i,{\rm D}}\cdot {\bf S}_{i+1,{\rm B}}
                     + {\bf S}_{i,{\rm C}^{\prime}}\cdot {\bf S}_{i+1,{\rm A}^{\prime}}]
\nonumber \\
&+&J_{2} \sum_{i}  [{\bf S}_{i,{\rm A}}\cdot {\bf S}_{i,{\rm D}}
                        +{\bf S}_{i,{\rm A}^{\prime}}\cdot {\bf S}_{i,{\rm D}}
                        +{\bf S}_{i,{\rm D}}\cdot {\bf S}_{i+1,{\rm A}}
                        +{\bf S}_{i,{\rm D}}\cdot {\bf S}_{i+1,{\rm A}^{\prime}}
],
\end{eqnarray}
where ${\bf S}_{i, \xi}$ is an $S=1/2$ spin operator at $\xi$-sublattice site in  $i$-th unit cell. 
Positions of the six sublattices in a unit cell are denoted by A, ${\rm A}^{\prime}$, B, C, ${\rm C}^{\prime}$ and D 
in Fig. \ref{fig1}(b).  
We fixed $J_{1}=1$  hereafter as a energy scale. 
In what follows, we examine the region of $0 \leq J_{2}/J_{1} \leq  3.5$ in the present study.
Note that the number of total spin sites is denoted by $N$; thus, the number of unit cells is $N/6$. 
\newline

 Let us introduce here the ground-state phase diagram of the $S=1/2$ Heisenberg model on the original Union Jack lattice of Fig. \ref{fig1}(a).
 At small $J_{2}/J_{1}$, one can see immediately that the antiferromagnetic N$\rm{\acute e}$el order is observed since this model 
corresponds to the $S=1/2$ antiferromagnetic Heisenberg one on the simple square lattice in the limit of $J_{2}/J_{1}=0$.
When the $J_{2}/J_{1}$ is increased, the intermediate-magnetization state appears.
A variational analysis for the classical model revealed that the spin configurations in this intermediate region are defined by the angle of cant $\varphi$ as illustrated in Fig. \ref{fig2}\cite{UJack4}.
Therefore, this intermediate-state is called canted-ferrimagnetic one.
The phase transition between the N$\rm{\acute e}$el and canted-ferrimagnetic phases occurs at $\alpha_{1} \equiv J_{2}/J_{1} \sim 0.84$ 
from the view point of the spin-wave theory\cite{UJack1, UJack2}. 
On the other hand, it was reported that this phase transition occurs at  $\alpha_{1} \sim 0.65$ by using the series expansion (SE)\cite{UJack3} and coupled cluster method (CCM)\cite{UJack4} techniques.
It was also discussed the possibility that the semistriped-ferrimagnetic state as illustrated in Fig. \ref{fig2} appears at very large values of $J_{2}/J_{1}$ ($\alpha_{2} \equiv J_{2}/J_{1} \approx 125$)\cite{UJack4}.

In what follows, we examine the ground-state phase diagram of the $S=1/2$ Heisenberg model on the Union Jack strip lattice depicted in Fig. \ref{fig1}(b) and compare the results of the original 2D lattice with those of the our strip lattice.

\section{Numerical methods}
We employ two reliable numerical methods: the exact diagonalization (ED) method and the density matrix renormalization group (DMRG) method\cite{DMRG1, DMRG2}. 
The ED method can
be used to obtain precise physical quantities for finite-size clusters.
This method does not suffer from the limitation of the shape of the clusters. 
It is applicable even to systems with frustration, in contrast to the quantum Monte Carlo (QMC) method 
coming across the so-called negative-sign problem for systems with frustration. 
The disadvantage of the ED method is the limitation that available system sizes are very small. 
Thus, we should pay careful attention to finite-size effects in quantities obtained from this method. 

On the other hand, the DMRG method is very powerful when a system is (quasi-)one-dimensional 
under the open-boundary condition. 
The method can treat much larger systems than the ED method. 
Note that the applicability of the DMRG method is irrespective of whether or not systems include frustrations. 
In the present research, we use the ``finite-system'' DMRG method. 
Note that we carefully choose the maximum number of retained states ($MS$) and the number of sweeps ($SW$) in our DMRG calculations.

\section{results}
\begin{figure}
\includegraphics[width=16.0cm]{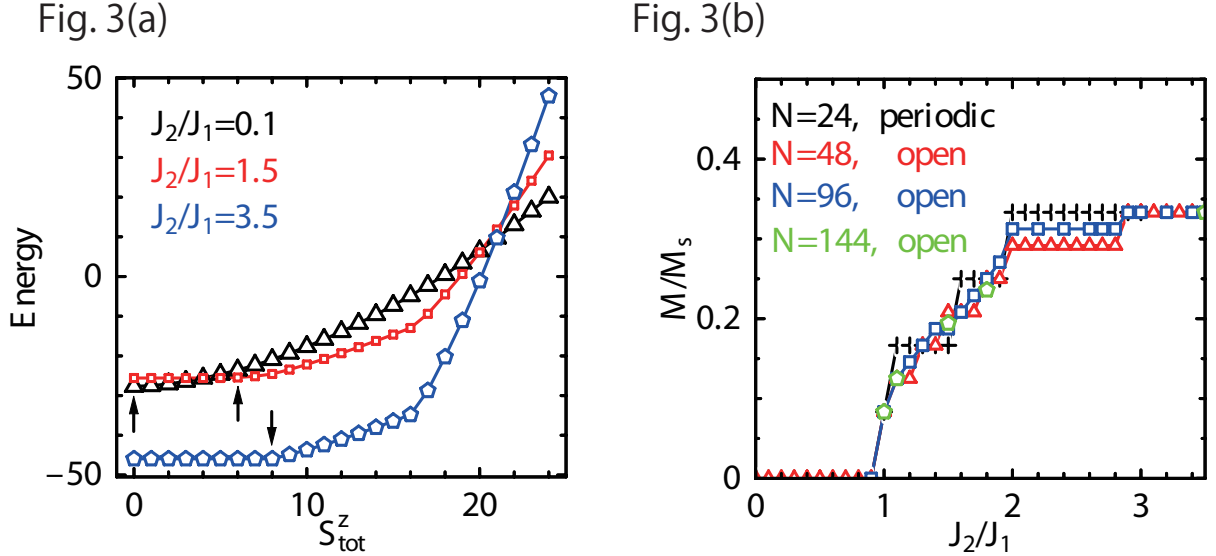}
\caption{(Color online) (a) Dependences of the lowest energy on $S_{\rm tot}^{z}$. 
Results of $J_{2}/J_{1}$ = 0.1,  1.5 and 3.5 for the system size of $N=48$ are presented. 
Arrows indicate the values of the spontaneous magnetization $M$ in each $J_{2}/J_{1}$.
(b)$J_{2}/J_{1}$-dependence of $M/M_{\rm s}$ obtained from ED calculations 
for $N=24$ (black cross) under the periodic-boundary condition 
and DMRG calculations for $N=48$ (red triangle), $96$ (blue square) and $144$ (green pentagon) 
under the open-boundary condition.
}\label{fig3}
\end{figure}

In this section, we present our numerical results in the ground-state of the $S=1/2$ Heisenberg model on the Union Jack strip lattice of Fig. \ref{fig1}(b).
First, let us explain the way to obtain the spontaneous magnetization $M$ in the ground state of the quantum system with isotropic interactions.
We calculate the lowest energy $E(J_{2}/J_{1}$, $S_{\rm tot}^{z}$, $N)$, 
where $S_{\rm tot}^{z}$ is the $z$-component of the total spin. 
For example, the energies for each $S_{\rm tot}^{z}$ in the three cases of $J_{2}/J_{1}$ are shown in Fig. \ref{fig3}(a).
In this figure, the results of the DMRG calculations with the $MS=700$ and $SW=15$ are presented when the systems size is $N=48$ for $J_{2}/J_{1}$=0.1, 1.5, 3.5.
The spontaneous magnetization $M(J_{2}/J_{1}$, $N$) is determined as the highest $S_{\rm tot}^{z}$ 
among those at the lowest common energy [see arrows in Fig. \ref{fig3}(a)]. 

\begin{figure}
\includegraphics[width=16.0cm]{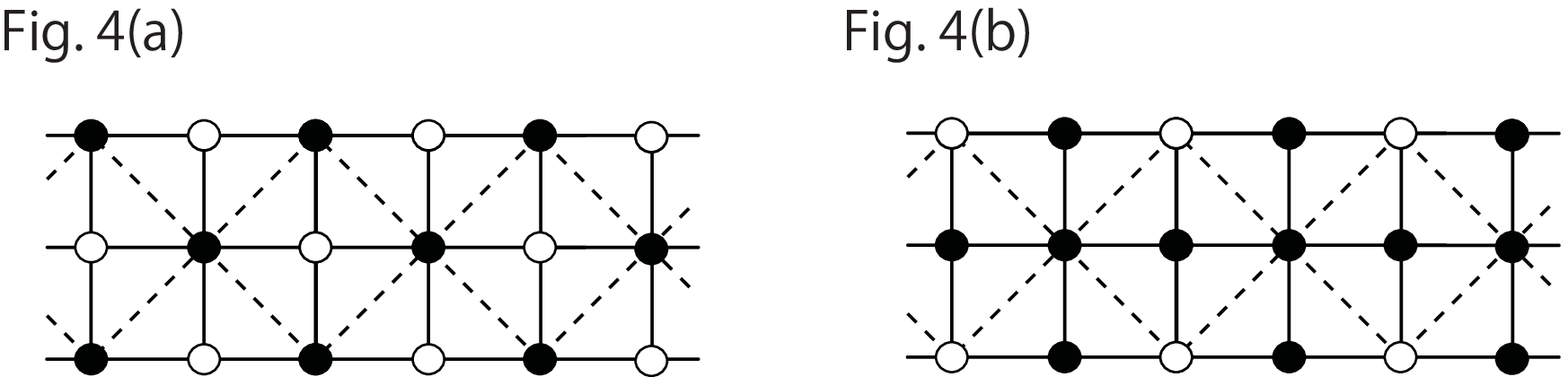}
\caption{(Color online) (a) Spin configuration in the  N$\rm{\acute e}$el-like spin liquid phase.
(b) Spin configuration in the LM ferrimagnetic phase of $M/M_{\rm s}=1/3$, where this configuration is obtained from the numerical results of the local magnetization shown in Fig. \ref{fig5}(b).
}\label{fig4}
\end{figure}

Our results of the $J_{2}/J_{1}$-dependence of the $M/M_{\rm s}$ are shown in Fig. \ref{fig3}(b), where $M_{\rm s}$ means saturated magnetization value, namely, $M_{\rm s}=N/2$.
In the limit of $J_{2}/J_{1}=0$, this Union Jack strip model depicted in Fig. \ref{fig1}(b) is reduced to the $S=1/2$ antiferromagnetic Heisenberg model on the three-leg ladder as known the typical system of the gappless spin-liquid ground states\cite{Dagotto, 3LegLadder1}.  
According to the study of the $S=1/2$ frustrated three-leg spin ladder\cite{TriLadder}, it is expected that the N$\rm{\acute e}$el-like spin liquid phase occurs in the ground-state when the strength of the $J_{2}/J_{1}$ is small but finite, where the schematic spin configuration in the N$\rm{\acute e}$el-like spin liquid state is illustrated in Fig. \ref{fig4}(a).
Indeed, our numerical calculations lead to the same conclusions that the nonmagnetic phase of $M/M_{\rm s}=0$ appears in the region where $J_{2}/J_{1}$ is relatively small.

For larger $J_{2}/J_{1}$, on the other hand, the magnetic phases with $M/M_{\rm s} \not = 0$ appears in the ground state.
Careful observation enables us to find that there are two magnetic phases in the thermodynamic limit; one is the intermediate magnetic phase of $0<M/M_{\rm s}<1/3$ 
and the other is the phase of $M/M_{\rm s}=1/3$. 
It should be noted here that the phase of $M/M_{\rm s}=(\frac{1}{3}-\frac{2}{N})$ which is found only under the open-boundary condition merges with the phase of $M/M_{\rm s}=1/3$ in the thermodynamic limit of $N \rightarrow \infty$ since the value of $M/M_{\rm s}$ becomes gradually larger and approaches the value of $M/M_{\rm s}=1/3$. This change due to the increase of the system size comes from finite-size effect. 
It is important that we successfully observe the intermediate-magnetization phase where the spontaneous magnetization $M/M_{\rm s}$ changes continuously with respect 
to the strength of $J_{2}/J_{1}$.

\begin{figure}
\includegraphics[width=16.0cm]{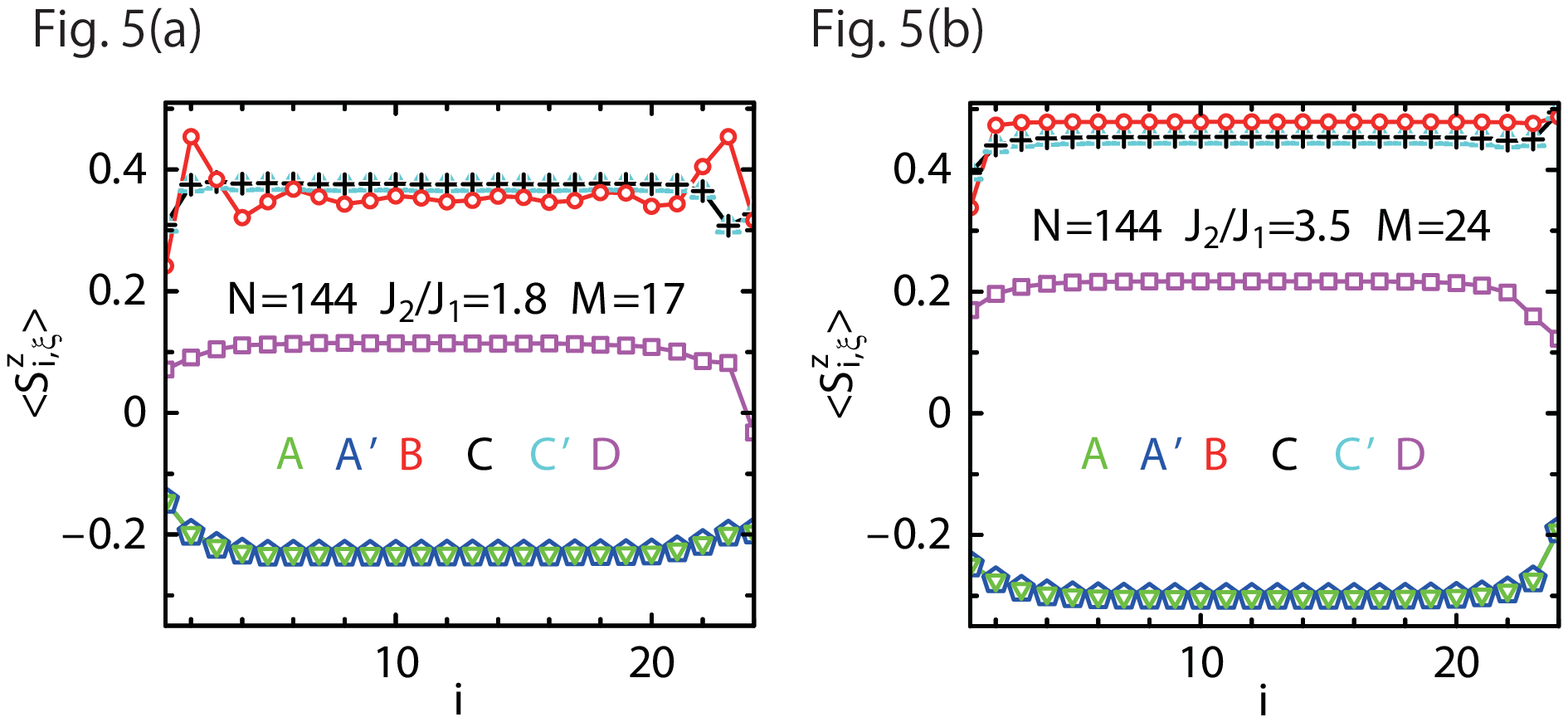}
\caption{(Color online) Local magnetization  $\langle S_{i,\xi }^{z} \rangle$ at each sublattice $\xi$.
Panels (a) and (b) are results for $J_{2}/J_{1}=$1.8 and 3.5 respectively.
These results are obtained from our DMRG calculations for $N=144$ ($i=$1,2, $\cdots$, 24).
}\label{fig5}
\end{figure}

Next, we calculate the local magnetization $\langle S_{i, \xi}^{z} \rangle$ to investigate the spin configurations in these two magnetic states, 
where $\langle A \rangle$ denotes the expectation value of the physical quantity $A$ and $S_{i, \xi}^{z}$ is the $z$-component of ${\bf S}_{i, \xi}$. 
Figure \ref{fig5} depicts our results for a system size $N=144$ on the lattice depicted in Fig. \ref{fig1}(b) under the open-boundary condition; 
Fig. \ref{fig5}(a) and Fig. \ref{fig5}(b) correspond to the case of $J_{2}/J_{1}=$1.8 and 3.5 respectively; we use green inverted triangle for $\xi=$A, 
blue pentagon for $\xi={\rm A}^{\prime}$, 
red circle for $\xi={\rm B}$, 
black cross for $\xi=$C, 
aqua triangle for $\xi={\rm C}^{\prime}$, 
and purple square for $\xi=$D. 
In Fig. \ref{fig5}(a), we find clearly incommensurate modulations with long-distance periodicity in the behavior of the local magnetization at the $B$-sublattice sites.
Therefore, we conclude that the intermediate-magnetization phase of $0<M/M_{\rm s}<1/3$ is the NLM ferrimagnetic one.
On the other hand, in Fig. \ref{fig5}(b), we observe the uniform behavior of upward-direction spins at sublattice-sites B, C, ${\rm C}^{\prime}$ and D and downward-direction spins at sublattice A and ${\rm A}^{\prime}$ as illustrated in Fig. \ref{fig4}(b).
The spin configuration in the phase of $M/M_{\rm s}=1/3$ can be understood from the viewpoint of the MLM theorem because the present strip model corresponds to the $S=1/2$ antiferromagnetic Heisenberg model on the diamond chain in the limit of $J_{2}/J_{1}=\infty$. 
Therefore, it is naturally lead to the conclusion that the phase of $M/M_{\rm s}=1/3$ is the LM ferrimagnetic one.

Finally, we discuss the relationship between the ground-state properties of the original Union Jack model depicted Fig. \ref{fig1}(a) and those of the strip model depicted in Fig. \ref{fig1}(b).
It is confirmed that the schematic spin configuration in the N$\rm{\acute e}$el-like spin liquid state depicted in Fig. \ref{fig4}(a) is consistent to 
that in the N$\rm{\acute e}$el state depicted in Fig. \ref{fig2}. 
We also confirm that the schematic spin configuration depicted in Fig. \ref{fig4}(b) agrees completely with that in the semistriped-ferrimagnetic state of the original Union Jack model as shown in Fig. \ref{fig2}.  
In addition, there exists the intermediate-magnetization state where the spontaneous magnetization is gradually changed in the ground-state of the both models although the incommensurate modulation with long-distance periodicity has not been confirmed in the case of 2D Union Jack model.
Therefore,  one finds that the the ground-state phase diagram of the $S=1/2$ antiferromagnetic Heisenberg model on the Union Jack strip lattice is qualitatively consistent to that of the $S=1/2$ antiferromagnetic Heisenberg model on the original Union Jack lattice.
The intermediate-magnetization state of the original Union Jack lattice have been understood from the view point of the classical configuration, namely "canted-state".
However, our numerical results of the Union Jack strip model leads to the possibility that the intermediate-state of the original Union Jack lattice is also the NLM ferrimagnetic one 
whose characteristic behavior in the local magnetization originates from pure quantum effects.
The future studies are desirable to confirm the presence of the incommensurate modulation in the canted-state of the 2D Union Jack model.

\section{CONCLUSIONS}
We have studied the ground-state properties of the $S=1/2$ antiferromagnetic Heisenberg model on the Union Jack strip lattice depicted in Fig. \ref{fig1}(b)  
by the ED and DMRG methods. 
Our numerical calculations have revealed that the intermediate-magnetization state occurs between the N$\rm{\acute e}$el-like spin liquid state corresponding to the 
N$\rm{\acute e}$el state of the original 2D model and the LM ferrimagnetic state which agrees with the semistriped-ferrimagnetic state of the original 2D model.
In this intermediate-magnetization state of this strip model, the spontaneous magnetization changes gradually with respect to the strength of the inner interaction.
We have also found the existence of the incommensurate modulation with long-distance periodicity of the local magnetization. 
From the finds of these characteristic behavior, it has concluded that the intermediate state of this strip model is the NLM ferrimagnetism.
These results naturally lead to the expectation that the intermediate-magnetization state of the original model is also the NLM ferrimagnetic one.

\begin{acknowledgments}
We wish to thank Professor T. Sakai for fruitful discussions. 
One of the authors (T. S.) acknowledges the financial support from the Motizuki Fund of Yukawa Memorial Foundation.
This work was partly supported  
by Grants-in-Aid 
(Nos. 20340096, 23340109, 23540388, and 24540348) 
from the Ministry of Education, Culture, Sports, 
Science and Technology of Japan.
Diagonalization calculations in the present work were 
carried out based on TITPACK Version 2 coded by H. Nishimori.
DMRG calculations were carried out 
using the ALPS DMRG application\cite{ALPS}.
Part of the computations were performed 
using the facilities of 
the Supercomputer Center, Institute for Solid 
State Physics, University of Tokyo.

\end{acknowledgments}

\end{document}